\documentclass[structabstract]{aa}  
\usepackage{graphicx}
\usepackage{natbib}
\usepackage{txfonts}
\begin{document}
   \title{Impact of rotation and disc lifetime on pre-main sequence lithium depletion of solar-type stars}
   \titlerunning{Rotation, disc lifetime and pre-main sequence lithium depletion}

   \author{P. Eggenberger \and L. Haemmerl\'{e} \and G. Meynet \and A. Maeder}

\institute{Observatoire de Gen\`eve, Universit\'e de Gen\`eve, 51 Ch. des Maillettes, CH-1290 Sauverny, Suisse\\ 
	\email{[patrick.eggenberger;lionel.haemmerle;georges.meynet;andre.maeder]@unige.ch}
             }


 
  \abstract
   {}
   {We study the influence of rotation and disc lifetime on lithium depletion of pre-main sequence (PMS) solar-type stars.}
   {The impact of rotational mixing and of the hydrostatic effects of rotation on lithium abundances are
investigated by computing non-rotating and rotating PMS models that include a comprehensive treatment of shellular rotation.
The influence of the disc lifetime is then studied by comparing the lithium content 
of PMS rotating models experiencing different durations of the disc-locking phase between 3 and 9\,Myr.}
   {The surface lithium abundance at the end of the PMS is decreased when rotational effects are included. 
During the beginning of the lithium depletion phase, only hydrostatic effects of rotation are at work. This results in a decrease
in the lithium depletion rate for rotating models compared to non-rotating ones. When the convective envelope recedes 
from the stellar centre, rotational mixing begins to play an important role due to differential rotation near the
bottom of the convective envelope. This mixing results in a decrease in the surface lithium abundance
with a limited contribution from hydrostatic effects of rotation, which favours lithium depletion during
the second part of the PMS evolution.
The impact of rotation on PMS lithium depletion is also found to be sensitive to the duration of the
disc-locking phase. When the disc lifetime increases, the PMS lithium abundance of a solar-type star
decreases owing to the higher efficiency of rotational mixing in the radiative zone. 
A relationship between the surface rotation and lithium abundance at the end of the PMS is then obtained: 
slow rotators on the zero-age main sequence are predicted to be more lithium-depleted 
than fast rotators due to the increase in the disc lifetime.}
   {}


   \maketitle
%

\section{Introduction}

Rotation is one of the key processes to influence all outputs of stellar models \citep[see e.g.][]{mae09}.
While rotational effects have an especially strong impact on the physics and evolution of massive stars, 
they can also change the global properties and evolution
of solar-type stars by counteracting the effets of atomic diffusion and bringing fresh hydrogen fuel to the central stellar core
\cite[e.g.][]{egg10_melange}. 
Observations of light element abundances bring valuable constraints to progress in our modelling
of these transport processes \citep[e.g.][]{pin10}. In a preceding work \citep{egg10_exo}, we investigated the influence 
of shellular rotation on the lithium abundances of main-sequence solar-type stars in the context of
a possible link between lithium depletion and the rotational history of exoplanet-host stars proposed by \cite{bou08}.
We then found that the larger efficiency of rotational mixing
predicted in exoplanet-host stars can result in lower lithium abundances for these
stars during the main sequence. 
After this study of the impact of shellular rotation on the lithium abundances of main-sequence stars,
we are now interested in investigating the effects of rotation on the lithium content of 
pre-main sequence (PMS) solar-type stars.

The impact of rotation on PMS lithium depletion of solar-type stars 
has already been discussed \cite[e.g.][]{pin90,mar96,men99}. 
These previous studies focused mainly on the hydrostatic effects of rotation\footnote{By hydrostatic effects of
rotation, we mean here the effects related to the change in the effective gravity due to rotation (the effective gravity
being defined as the classical gravity diminished by the centrifugal acceleration).}
and did not consider the influence of a possible coupling between the star and its disc.
The star-disc interaction plays a key role in determining the rotation of the star on the zero-age main sequence (ZAMS).
By assuming that the presence of the disc prevents the star from spinning up so that its 
surface angular velocity remains constant, one finds that the influence of this coupling simply relies on 
the disc lifetime. A longer disc lifetime then enables the star to lose a larger amount of angular momentum during the PMS, 
resulting in lower rotation rates on the ZAMS. 
In addition to its effect on the surface rotational velocity during the PMS, the duration of the disc-locking phase 
can change the internal rotation profiles, with
a possible impact on the lithium content. In this paper, we thus investigate 
the influence of rotation and disc lifetime on the PMS lithium depletion of solar-type stars.   
Both effects of rotational mixing and of the centrifugal force are
studied by computing models that include a comprehensive treatment of shellular rotation.
The effects of the disc lifetime on the PMS lithium depletion are then investigated by comparing the
PMS evolution of models with different durations of the disc-locking phase. 

The modelling of rotation is first briefly described in Sect.~\ref{models}. The effects of rotation
on PMS lithium depletion are discussed in Sect.~\ref{res_rot}, while the impact of the disc lifetime is
investigated in Sect.~\ref{res_disc}. The conclusion is given in Sect.~\ref{conclusion}.

\section{Modelling rotation}
\label{models}

The stellar evolution code used for these computations is the Geneva code,
which includes a detailed treatment of shellular rotation \citep{egg08}.
In the context of shellular rotation, the transport of angular momentum in radiative zones obeys an advection-diffusion equation written in Lagrangian
coordinates \citep{zah92, mae98}:
\begin{equation}
  \rho \frac{{\rm d}}{{\rm d}t} \left( r^{2}\Omega \right)_{M_r} 
  =  \frac{1}{5r^{2}}\frac{\partial }{\partial r} \left(\rho r^{4}\Omega
  U(r)\right)
  + \frac{1}{r^{2}}\frac{\partial }{\partial r}\left(\rho D r^{4}
  \frac{\partial \Omega}{\partial r} \right) \, , 
\label{transmom}
\end{equation}
where $r$ is a characteristic radius, $\rho$ the mean density on an isobar, and $\Omega(r)$ the mean angular velocity at level $r$.
The vertical component $u(r,\theta)$ of the velocity of the meridional circulation at a distance
$r$ to the centre and at a colatitude $\theta$ can be written
\begin{equation}
u(r,\theta)=U(r)P_2(\cos \theta)\,.
\end{equation}
Only the radial term $U(r)$ appears in Eq. (\ref{transmom}).
Its expression is given by \citep{mae98}
\begin{eqnarray}
U(r)  =  \frac{P}{\rho g C_{P} T [\nabla_{\rm ad}-\nabla + (\varphi/\delta)
  \nabla_{\mu}]}
 \cdot  \left\{ \frac{L}{M}(E_{\Omega }+E_{\mu}) \right\}\,, 
\label{vmer}
\end{eqnarray}
where $P$ is the pressure, $C_P$ the specific heat, $E_{\Omega}$ and $E_{\mu}$ are terms
depending on the $\Omega$- and $\mu$-distributions respectively, up to the third-order 
derivatives and on various thermodynamic quantities \cite[see][for more details]{mae98}. 

Meridional circulation and shear mixing are considered as the main mixing mechanisms in radiative zones.
In convective zones, solid-body rotation is assumed as indicated by the solar case.
The first term on the right-hand side of Eq. (\ref{transmom}) describes the advection of angular momentum by 
meridional circulation, while the second term accounts for the transport of angular momentum by shear instability
with $D=D_{\rm shear}$. The expression of this diffusion coefficient is given by
\begin{eqnarray}
D_{\rm shear} & = & \frac{ 4(K + D_{\mathrm{h}})}
{\left[\frac{\varphi}{\delta} 
\nabla_{\mu}(1+\frac{K}{D_{\mathrm{h}}})+ (\nabla_{\mathrm{ad}}
-\nabla_{\mathrm{rad}}) \right] } \nonumber\\
&  & \times \frac{H_{\mathrm{p}}}{g \delta} \; 
 f_{\rm energ}\left( \frac{9 \pi}{32} \, \Omega{{\rm d}\ln \Omega \over {\rm d}\ln r} \right)^2 \, ,
\label{dshear}
\end{eqnarray}
with $K$ the thermal diffusivity, $f_{\rm energ}$ the fraction of the
excess energy in the shear that contributes to mixing (here taken equal to 1) \citep{mae01}, and 
$D_{\rm h}$ the diffusion coefficient associated
to horizontal turbulence. The usual expression for this coefficient is, according to \cite{zah92},
\begin{equation}
D_{\rm h} = \frac{1}{c_{\rm h}} r |2V(r)-\alpha U(r)|\,,
\label{Dhzahn}
\end{equation}
where $U(r)$ is the vertical component of the meridional circulation velocity, $V(r)$ the
horizontal component, $c_{\rm h}$ a constant of order unity (here taken equal to 1) and 
$\alpha=\frac{1}{2} \frac{{\rm d} \ln r^2 {\bar \Omega}}{{\rm d} \ln r}$. 
The full solution of Eq. (\ref{transmom}) taking $U(r)$ and $D$ into account gives the non-stationary solution of the problem. 
We recall here that meridional circulation is treated as a truly advective
process in the Geneva evolution code.

The vertical transport of chemicals
through the combined action of vertical advection and strong
horizontal diffusion 
can be described as a pure diffusive process \citep{cha92}.
The advective transport is thus replaced by a diffusive term, 
with an effective
diffusion coefficient
\begin{equation}
D_{\rm eff} = \frac{|rU(r)|^2}{30D_{\rm h}}\,.
\label{Deff}
\end{equation} 
The vertical transport of chemical elements then
obeys a diffusion equation that, in addition to this macroscopic transport,
also accounts for (vertical) turbulent transport:
\begin{eqnarray}
 \frac{{\rm d} X_i}{{\rm d} t} = \dot{X_i}
 + \frac{1}{\rho r^2}\frac{\partial}{\partial r}\left[ r^2\rho
   (D_{\rm eff}+D_{\rm shear})\frac{\partial X_i}{\partial r} \right]\,,
\label{transchem}
\end{eqnarray}
where $X_i$ is the mass fraction of element $i$, and $\dot{X_i}$ represents the variations in chemical
composition due to nuclear reactions.

\section{Effects of rotation}
\label{res_rot}

To study the effects of rotation on the lithium abundances of PMS solar-type stars,
we compute the PMS evolution of 1\,$M_{\odot}$ models with a solar chemical composition 
as given by \cite{gre93} and a solar-calibrated value for the mixing-length parameter. 
One model is first computed without rotation. The corresponding rotating model 
shares exactly the same initial parameters except for including a comprehensive treatment of shellular rotation. 
This model is computed with an initial angular velocity of 20\,$\Omega_\odot$ and a disc lifetime of 6\,Myr. During the disc-locking phase,
the surface angular velocity of the star is simply assumed to remain constant. 

\begin{figure}
\resizebox{\hsize}{!}{\includegraphics{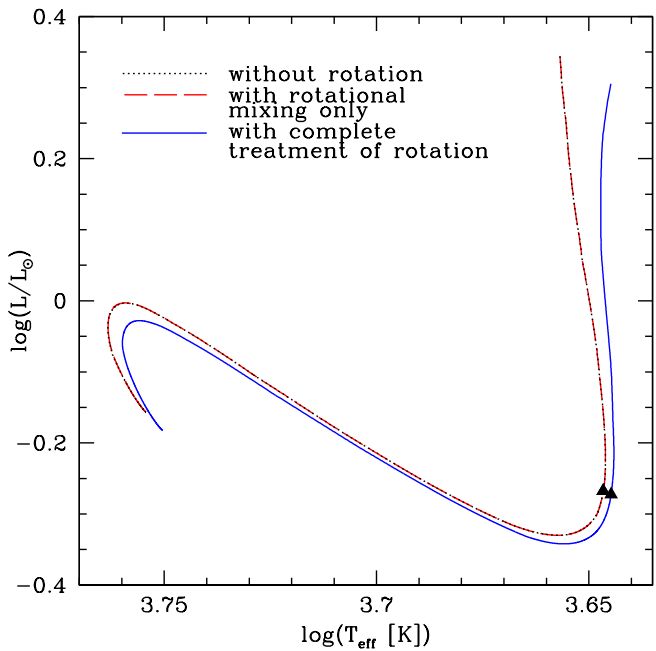}}
\caption{Effects of rotation on the PMS evolutionary tracks in the HR diagram for 1\,$M_{\odot}$ models. 
The dotted and continuous lines correspond to a non-rotating model and a rotating model computed with 
a full treatment of shellular rotation, respectively. The track corresponding to a rotating model including only
the effects of rotational mixing is shown by a dashed line and is superimposed on the track of the non-rotating model. 
The tracks end when the ZAMS is reached; the end of the disc-locking phase is indicated by a triangle.}
\label{dhr_rot}
\end{figure}

\begin{figure}
\resizebox{\hsize}{!}{\includegraphics{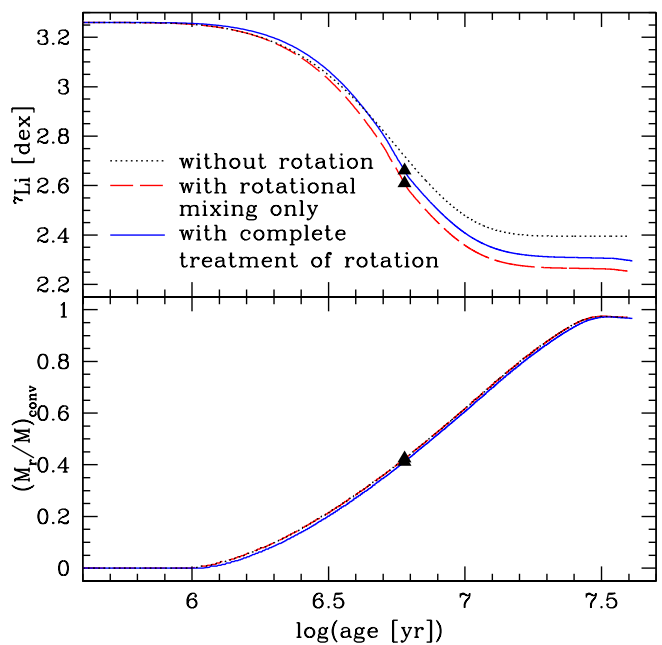}}
\caption{{\it Top}: effects of rotation on lithium depletion during the PMS for 1\,$M_{\odot}$ models. The non-rotating model is indicated by the dotted line, while the
dashed and continuous lines correspond to rotating models that only include the effects of rotational
mixing and a full treatment of shellular rotation, respectively. These rotating models are computed with the same initial 
angular velocity and disc lifetime. The tracks end when the ZAMS is reached and a triangle indicates
the end of the disc-locking phase. {\it Bottom}: location of the mass coordinate of the base of the convective envelope. The 
track corresponding to a rotating model that only includes
the effects of rotational mixing is superimposed on the track of the non-rotating model.}
\label{ab_lithium}
\end{figure}

The evolutionary tracks in the HR diagram are shown for both models in Fig.~\ref{dhr_rot}. 
Including rotation shifts the track to slightly lower effective temperatures and luminosities as previously found by many
studies \cite[see e.g.][]{pin89,mar96,men99}. 
Figure~\ref{ab_lithium} compares the evolution of the surface lithium abundance during the PMS for the models computed
with and without rotation starting with a lithium abundance 
$A({\rm Li})=\log [N({\rm Li})/N(H)]+12=3.26$  \citep[][]{asp09}. 
At the end of the PMS, the rotating model exhibits a lower surface lithium abundance than the non-rotating 
one (difference of about $0.1$\,dex for a total lithium depletion of about 1\,dex).
We thus see that the inclusion of rotation results in a global increase of the lithium depletion during the PMS. 

To investigate whether these differences in the lithium content 
between rotating and non-rotating models are mainly due to rotational mixing or to hydrostatic effects of rotation, 
another rotating model of 1\,$M_{\odot}$ is computed by only accounting for the effects of rotational mixing. The initial rotational velocity
and disc lifetime of this model are thus identical as for the rotating model including the full treatment of shellular rotation, but 
the effects of the centrifugal force on the stellar structure are not taken into account. The evolution of this model in the HR diagram
is also shown in Fig.~\ref{dhr_rot}, but the corresponding track is superimposed on the one of the non-rotating model. 
The change in the evolutionary tracks in the HR diagram induced by rotation for PMS solar-type stars 
is thus mainly due to the effects of the centrifugal force and not to a change in the global properties induced by
rotational mixing.

Concerning the evolution of the lithium content during the PMS, the situation is quite different from the changes
observed in the HR diagram. Figure~\ref{ab_lithium} indeed shows that the lithium depletion is greater for the rotating model
including only the effects of rotational mixing than for the non-rotating model. While the effects of
rotation on the PMS evolutionary tracks in the HR diagram are mainly due to the centrifugal force, rotational mixing is found to
play an important role in the increase of lithium depletion for rotating models during the PMS. 
As seen in Fig.~\ref{ab_lithium}, both rotating models exhibit lower lithium surface abundances
than the non-rotating model on the ZAMS. We also note that the lithium
depletion occurs earlier for the rotating model, when accounting only for the effects of rotational mixing, than for the model that includes the full
treatment of rotational effects. This is related to the hydrostatics effects of rotation. Rotational mixing has a negligible impact on the global stellar structure during the PMS, so that 
the temperature profile, the depth of the convective envelope, and the evolutionary tracks in the HR diagram are identical for the
non-rotating model and the rotating model computed without the hydrostatics effects of rotation. 
For the model including a
complete treatment of rotation, the temperature at the base of the convective envelope is lower during the initial phase when the star is
still fully convective. This is caused by
the effective gravity of the star being reduced by the centrifugal acceleration term, and 
this results in a lower lithium depletion 
rate. Including the centrifugal force therefore leads to a higher lithium content during the beginning of the lithium depletion phase on
the PMS.  

After this initial phase, a radiative core begins to develop at the centre of the star. The temperature at the base of the
convective envelope is slightly higher for the model that includes a complete 
treatment of rotation than for the non-rotating one during the end of the PMS evolution. 
After decreasing the lithium depletion rate
during the beginning of the PMS, the hydrostatic effects of rotation are thus found to favour lithium depletion during the second part of the PMS, in
good agreement with the study by \cite{men99}.
More importantly, rotational mixing begins to play a dominant role during this phase. As can be seen in Fig.~\ref{ab_lithium} by comparing
the non-rotating model to the model that includes rotational mixing only, the impact of rotational mixing on 
the surface lithium abundance begins to be visible after about 3\,Myr. By transporting lithium to deeper layers where it is 
easily destroyed, such mixing generates an increase in the lithium depletion. 
In the present PMS models, shear mixing plays the dominant role for lithium depletion with a negligible contribution from meridional
currents. The decrease in the surface lithium abundance by rotational mixing is thus directly related to 
the increase of differential rotation in the stellar interior during the PMS evolution. 

\begin{figure}
\resizebox{\hsize}{!}{\includegraphics{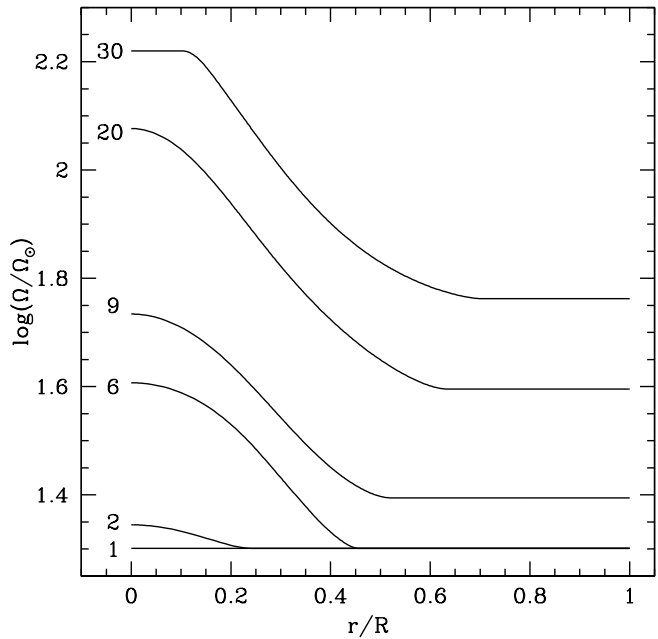}}
\caption{Evolution of the internal rotation profile for a 1\,$M_{\odot}$ rotating model with a disc lifetime of 6\,Myr. The age in Myr is
indicated for each rotation profile.}
\label{pro_rot}
\end{figure}

Figure~\ref{pro_rot} shows the internal rotation profile at different ages for the rotating model 
including a complete treatment of rotational effects. At the beginning of the PMS evolution, the star is fully convective
and rotates as a solid body. As soon as the radiative core develops, differential rotation takes place in the radiative zone mainly
as a result of the star contracting. During the disc lifetime, the star is assumed to be locked
to its disc with a constant surface angular velocity. As soon as the disc disappears, the angular velocity at the stellar surface
increases as shown in Fig.~\ref{pro_rot}. 
For the present model computed with a disc lifetime of 6\,Myr,
the angular velocity at the surface thus remains constant during the first 6\,Myr. Differential rotation
in the radiative zone is enhanced by disc locking and leads to a steep rotation profile at the base of the convective envelope. 
This generates turbulence below the convective envelope through shear instability, and the resulting mixing can then transport
lithium to deeper and hotter regions, where it is destroyed more efficiently.

\begin{figure}
\resizebox{\hsize}{!}{\includegraphics{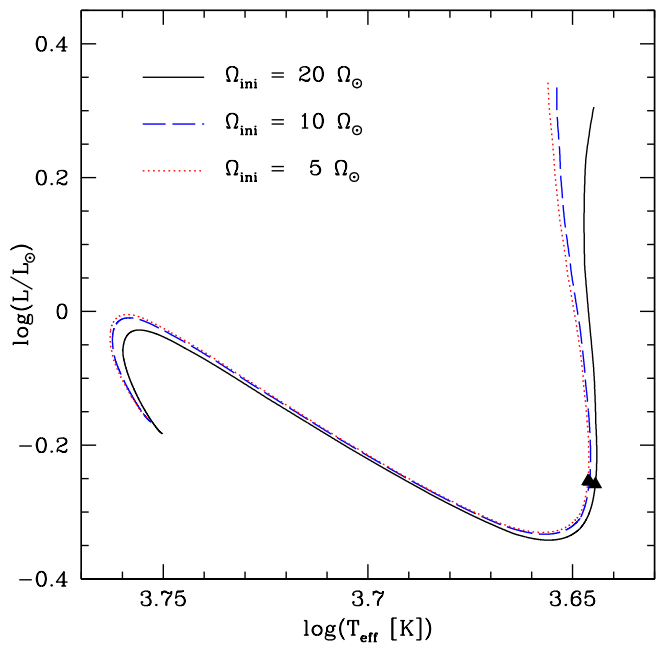}}
\caption{Evolutionary tracks in the HR diagram for rotating PMS models of 1\,$M_{\odot}$. The continuous, dashed, and 
dotted lines correspond to initial velocities of 20\,$\Omega_{\odot}$, 10\,$\Omega_{\odot}$, and 5\,$\Omega_{\odot}$, respectively. 
The tracks end when the ZAMS is reached and the end of the disc-locking phase is indicated by a triangle.}
\label{dhr_om_ini}
\end{figure}

\begin{figure}
\resizebox{\hsize}{!}{\includegraphics{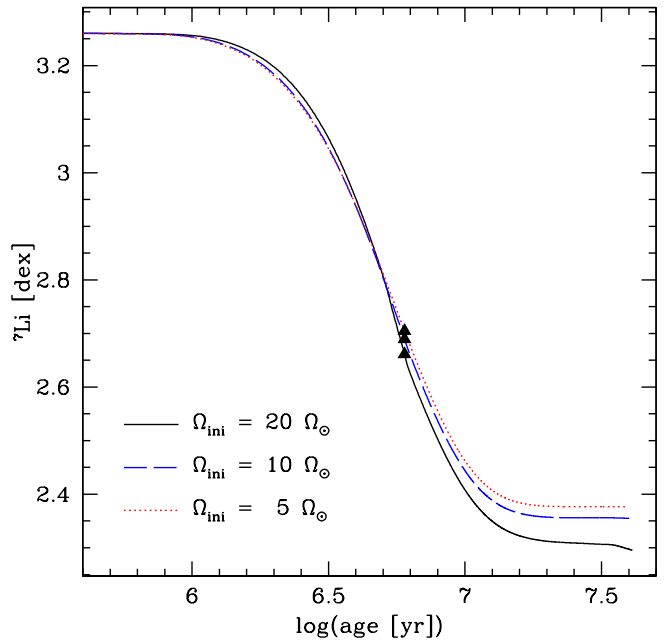}}
\caption{Lithium surface abundance during the PMS for rotating 1\,$M_{\odot}$ models. The continuous, dashed, and 
dotted lines correspond to initial velocities of 20\,$\Omega_{\odot}$, 10\,$\Omega_{\odot}$, and 5\,$\Omega_{\odot}$, respectively.
The triangles indicate the end of the disc-locking phase. The tracks end when the ZAMS is reached.}
\label{li_om_ini}
\end{figure}

The quantitative impact of rotation on the PMS lithium content of solar-type stars is of course sensitive to the initial angular velocity
adopted when the star is fully convective. To illustrate this point, two additional models of 1\,$M_{\odot}$ are computed 
with initial angular velocities of 5 and 10\,$\Omega_{\odot}$. These models are also computed with a solar chemical composition, 
a solar-calibrated value for the mixing-length parameter and a disc lifetime of 6\,Myr. Figure~\ref{dhr_om_ini} shows the
evolutionary tracks in the HR diagram of these models together with the 20\,$\Omega_{\odot}$ model discussed before. The evolution of 
the surface lithium abundance of these models is given in Fig.~\ref{li_om_ini}. The decrease in the lithium abundance at the end of the PMS
is found to be more pronounced for higher initial velocities. This mainly results from the increased efficiency of rotational mixing when the rotation rates increase. The hydrostatic effects of rotation are also increased when the
initial angular velocity is higher. As shown in Fig.~\ref{dhr_om_ini}, the shift to lower temperatures and luminosities 
becomes more visible in the HR diagram when the initial rotational velocity increases. This leads
to a slight decrease in the lithium depletion rate at the beginning of the PMS evolution when the initial velocity
increases. As seen in Fig.~\ref{li_om_ini}, lithium depletion therefore occurs earlier for lower initial rotation
velocities.   

\section{Effects of disc lifetime}
\label{res_disc}

In the preceding section, the effects of rotation on the PMS lithium depletion of solar-type stars have been discussed by
comparing rotating models computed with the same disc lifetime of 6\,Myr.
We are now interested in investigating the impact of the disc lifetime on the lithium depletion
during the PMS. Rotating models of 1\,$M_\odot$ are computed for different disc lifetimes of 3, 4, 5, 6, 7, 8, and 9\,Myr. 
All these models are computed with the same initial angular velocity of 20\,$\Omega_{\odot}$, a solar chemical composition, and
a solar-calibrated value for the mixing-length parameter. 
 
\begin{figure}
\resizebox{\hsize}{!}{\includegraphics{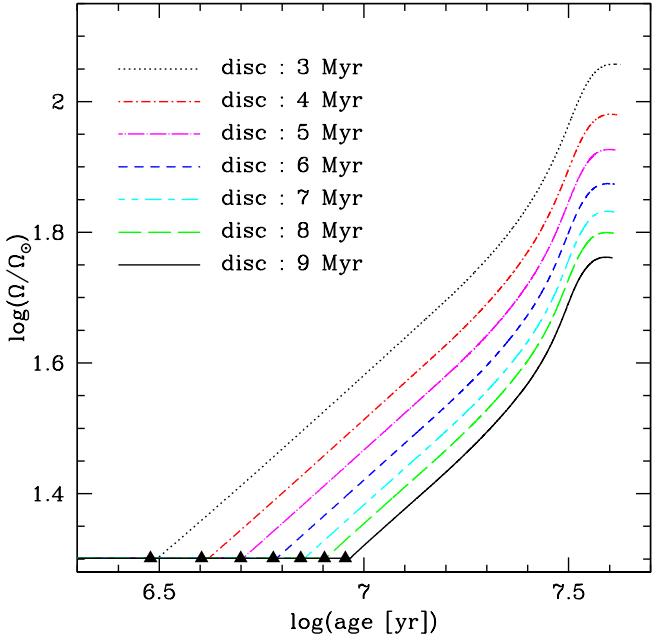}}
\caption{Evolution of the surface angular velocity for 1\,$M_{\odot}$ models sharing the
same initial velocity but computed with different disc lifetimes. Triangles indicate the end 
of the disc-locking phase for each model. The tracks end when the ZAMS is reached.}
\label{om_surf_td}
\end{figure}

The evolution of the surface angular velocity during the PMS is shown in Fig.~\ref{om_surf_td} for these models. The angular
velocity at the surface remains constant during the disc-locking phase. As soon as the disc disappears, the surface 
angular velocity increases as a result of the stellar contraction until the ZAMS is reached. When the disc lifetime increases, 
the star keeps its inital surface angular velocity for a longer time and loses a larger amount of angular momentum. As shown in
Fig.~\ref{om_surf_td}, the increased surface angular velocity then occurs later for longer disc lifetimes,  
resulting in lower rotational velocities on the ZAMS.

\begin{figure}
\resizebox{\hsize}{!}{\includegraphics{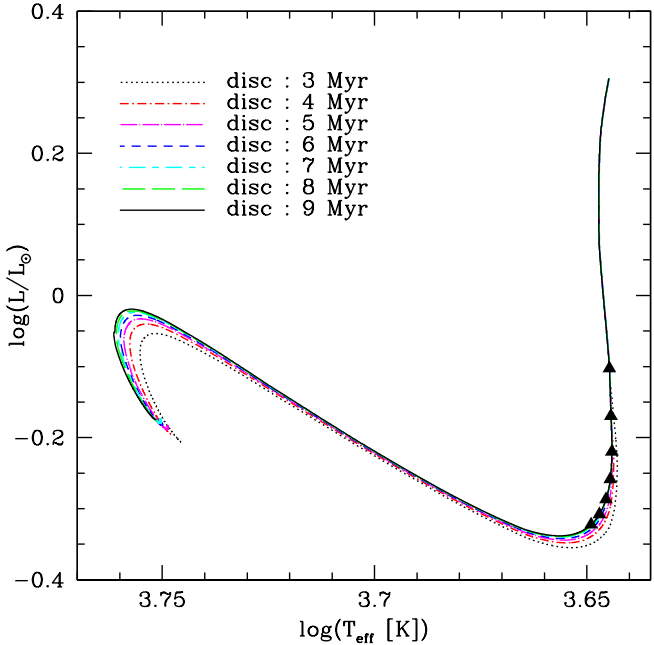}}
\caption{Evolutionary tracks in the HR diagram for 1\,$M_{\odot}$ models with the
same initial velocity but different disc lifetimes. Triangles indicate the end 
of the disc-locking phase for each model.}
\label{hr_td}
\end{figure}

Figure~\ref{hr_td} shows the evolutionary tracks in the HR diagram corresponding to these rotating 
models computed for various durations of the disc-locking phase. During the first 3\,Myr, all models follow the
same track in the HR diagram because they share the same rotational history. 
Rotating models with shorter
disc lifetimes exhibit higher rotational velocities as soon as the disc disappears, compared to rotating 
models that remain coupled to their disc for a longer time. Consequently, the hydrostatic effects of rotation are more
pronounced for shorter disc lifetimes; this results in larger shifts to the red part of the HR diagram when the disc lifetime
decreases.   

\begin{figure}
\resizebox{\hsize}{!}{\includegraphics{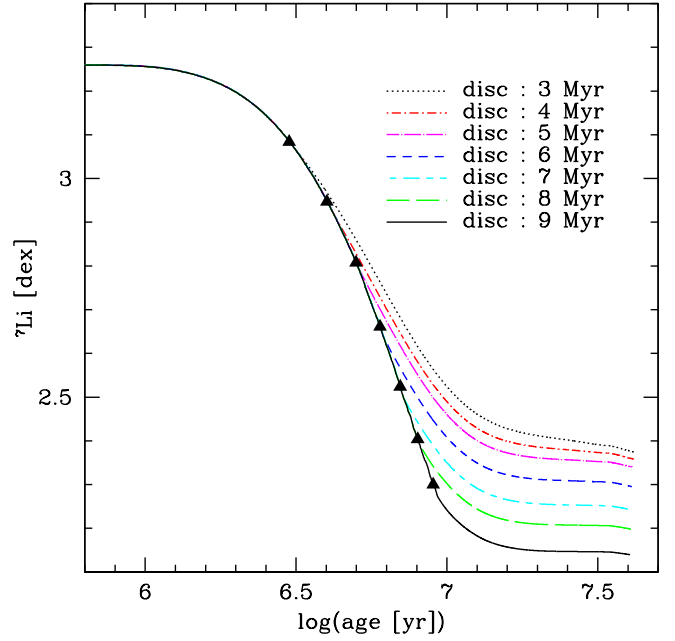}}
\caption{Evolution of the surface lithium abundance for 1\,$M_{\odot}$ models with the
same initial velocity but different disc lifetimes. Triangles indicate the end 
of the disc-locking phase for each model.}
\label{li_td}
\end{figure}

The evolution of the surface lithium abundance is shown in Fig.~\ref{li_td}. While the effects of rotation in the HR diagram
are found to be stronger for models with shorter disc lifetimes, we see exactly the opposite for the impact of disc lifetime
on lithium depletion. Figure~\ref{li_td} indeed shows that lithium depletion during the PMS increases for longer durations
of the disc-locking phase. Soon after the disappearance of the disc, the lithium depletion decreases compared to models with
longer disc lifetimes. As a result, models computed with shorter disc lifetimes exhibit higher lithium abundances on the ZAMS
than models experiencing a longer disc-locking phase.

\begin{figure}
\resizebox{\hsize}{!}{\includegraphics{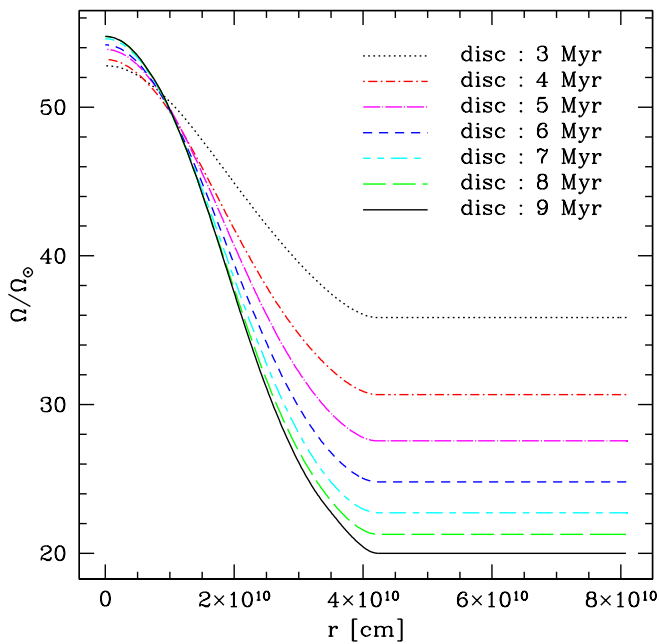}}
\caption{Rotation profiles at an age of 9\,Myr (corresponding to the disappearance of the disc for the model with the longer
disc lifetime) for models computed with different disc lifetimes.}
\label{prof_om_td}
\end{figure}

Figure~\ref{prof_om_td} shows the internal rotation profiles of these models with different disc lifetimes 
at an age of 9\,Myr, which corresponds to the end of the disc-locking phase for the model with the longer disc lifetime.
The gradient of angular velocity below the convective envelope increases when the disc lifetime increases. 
Rotational mixing is then more efficient in the interiors of models 
with longer disc lifetimes, which explains the increased lithium depletion shown in Fig.~\ref{li_td}. We thus conclude that an
increase of the disc lifetime leads to a decrease in the lithium content during the PMS evolution of a solar-type star owing to
the larger amount of differential rotation generated during the disc-locking phase. 

\begin{figure}
\resizebox{\hsize}{!}{\includegraphics{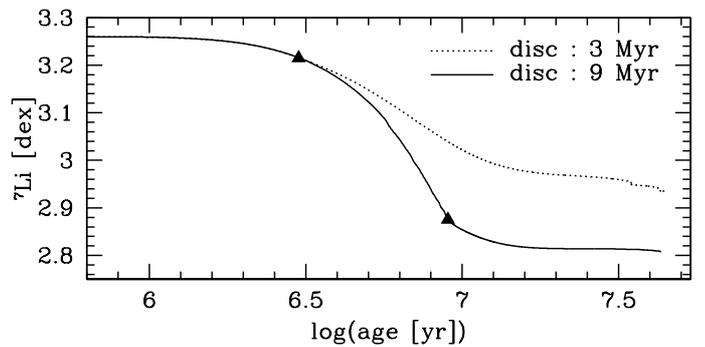}}
\caption{Evolution of the surface lithium abundance for two 1\,$M_{\odot}$ models computed with a
mixing-length parameter $\alpha=0.6\, \alpha_{\odot}$. The continuous and dotted lines correspond to disc lifetimes
of 9 and 3\,Myr, respectively. Triangles indicate the end 
of the disc-locking phase for each model.}
\label{li_td_alpha}
\end{figure}

The PMS lithium depletion predicted by 
non-rotating solar calibrated models is known to be too high to correctly reproduce the observations in open
clusters \citep[e.g.][]{dan94,ven98,sch99,pia02,dan03}.   
Since including rotational effects results in lower lithium abundances for rotating models compared 
to non-rotating ones (with a higher impact for longer disc lifetimes), rotation does not solve this problem.
One possibility for correctly reproducing the lithium abundances observed in open clusters is to reduce the value 
of the mixing-length parameter. 
The PMS evolution of two additional rotating models is then computed with a 
mixing-length parameter $\alpha=0.6\, \alpha_{\odot}$ and two different disc lifetimes of 3 and 9\,Myr.
Figure~\ref{li_td_alpha} shows the evolution of the surface lithium abundance for these models during the PMS.
As expected, the reduced value of the mixing-length parameter results in a lower PMS lithium depletion  
for both models. More importantly, we see that the increased duration of the disc-locking phase also leads
to a significant decrease in the lithium abundance on the ZAMS, in good agreement with the results obtained
above for solar-calibrated models.

\begin{figure}
\resizebox{\hsize}{!}{\includegraphics{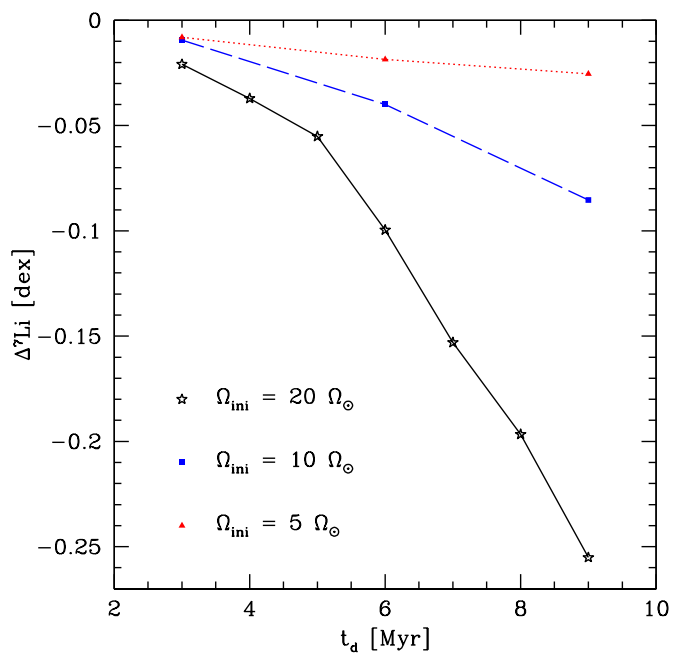}}
\caption{Surface lithium depletion on the ZAMS for 1\,$M_{\odot}$ models computed with 
different disc lifetimes (t$_{\rm d}$) and initial velocities ($\Omega_{\rm ini}$). $\Delta^7$Li is defined as the difference between the
surface lithium abundance on the ZAMS of rotating models and the non-rotating one.}
\label{zams}
\end{figure}

Two different parameters are found to
determine the quantitative impact of rotational effects on PMS lithium depletion: the initial rotational velocity 
and the disc lifetime. For a given disc lifetime but different initial velocities,
a direct correlation between the surface velocity and the
lithium abundance at the end of the PMS is then obtained so that fast rotators are predicted to be more lithium depleted than 
slow rotators. For a given initial velocity but different disc lifetimes, one expects that
slow rotators on the ZAMS are more lithium depleted than fast rotators. 
This is illustrated in Fig.~\ref{zams}, which shows the lithium depletion on the ZAMS as a function of the 
disc lifetime for rotating models computed with different initial velocities. We see that models with
longer disc lifetimes, hence lower rotation rates on the ZAMS, exhibit higher lithium depletions than
models with shorter disc lifetimes that are fast rotators on the ZAMS.
This agrees well with 
observations in the Pleiades of higher lithium abundances for fast rotators than for slow rotators \citep{sod93}. 
Moreover, longer disc lifetimes may favour the formation and migration of giant exoplanets, so that
lower lithium abundance for exoplanet-host stars are also predicted in this context. This agrees well with the
possible detection of differences in the lithium content of stars with and without exoplanets reported by \cite{isr09} 
\cite[but see also][for a contrary view]{bau10}.

We may wonder whether the 
results obtained here in the context of PMS models including shellular rotation are purely academic, given the need 
of another mechanism for the transport of angular momentum to account for the solar rotation profile for instance 
\cite[e.g.][]{mes87,cha93,egg05_mag,den10}. The rotational periods 
of young solar-type stars suggest that slow rotators develop a high degree of differential rotation between the radiative core 
and the convective envelope \citep[see e.g.][]{bou08,den10_etal}. This indicates that the hypothesis of shellular rotation during
the PMS evolution is perfectly valid in this case even if an additional process can eventually intervene 
later on during the main-sequence evolution, leading to a flat rotation profile at the solar age. 

For fast rotators, the situation is somewhat different. The observation of rotational periods of young solar-type stars 
suggests that an efficient mechanism for the transport of angular momentum is needed for fast rotators 
to correctly reproduce their spin-down in open clusters \citep[e.g.][]{bou08,den10_etal}. The 
hypothesis of shellular rotation during the PMS thus seems more questionable in the case of fast rotators than for 
slow rotators. This hypothesis is however not incompatible with the observation of rotational periods for fast
rotating solar-type stars, because the strong internal coupling suggested by these observations is not needed 
before the beginning of the main sequence. This means that a fast rotator can exhibit differential rotation during the PMS with
an additional mechanism for the transport of angular momentum at work with a short coupling timescale, leading to an approximate
flat rotation profile at the end of the PMS. Moreover, Fig.~\ref{li_td} shows that the effects of rotation and disc lifetime on
lithium depletion occur mainly during the first 10\,Myr of the PMS evolution. The results obtained in this study will then
remain qualitatively valid even in the case of a fast rotator that only exhibits differential rotation during the first 10\,Myr 
and then rotates approximately as a solid-body during the rest of its PMS evolution. 
We thus conclude that even if the physical mechanisms responsible for the transport of angular momentum in the interiors of solar-type stars are
currently not fully understood, the assumption of shellular rotation during the PMS seems to be reasonable according to current
observational constraints.

We note that including a mechanism
that enforces solid-body rotation during the whole PMS evolution will change the picture. 
In the case of solar-type stars, the efficiency of rotational mixing is indeed strongly reduced when solid-body rotation is imposed
 \cite[see for more details Sect. 3 of][]{egg10_melange}. The effects of rotational mixing then become negligible compared
to hydrostatic effects of rotation. As a result, the impact of disc lifetime on lithium depletion differs from the one described
above. For PMS models in solid-body rotation, a longer disc lifetime reduces the rotational velocity of the star, which simply results in a
decrease in the hydrostatic effects of rotation (without any increase in the efficiency of rotational mixing), hence to an overall
decrease in the impact of rotation on lithium depletion. The relationship between slow rotation and low lithium abundance 
on the ZAMS is then found to disappear when solid-body rotation is enforced during the PMS.

\section{Conclusions}
\label{conclusion}

The effects of rotation on the PMS surface lithium abundance of solar-type stars have been 
studied. Accounting for rotational effects results
in an overall decrease in the lithium content at the end of the PMS.
The effects of rotation on the PMS lithium content of solar-type stars are enhanced when the initial velocity of the
star increases. Moreover, the impact of rotation on PMS lithium depletion is found to depend on the duration of the
disc-locking phase, during which the surface angular velocity of the star remains constant. 
When the disc lifetime increases, the PMS lithium abundance of a solar-type star is indeed found to
decrease. This is due to the increase in the efficiency of rotational mixing, which is related 
to the larger amount of differential rotation 
generated in the radiative zone during the disc-locking phase. This illustrates 
that the efficiency of shear mixing is more
direcly related to differential rotation in stellar interiors than to surface rotational velocities, as also found in the
case of stars undergoing magnetic braking \cite[e.g.][]{mey11}.     

A link between the surface velocity and lithium abundance at the end of the PMS is then obtained: 
slow rotators on the ZAMS are predicted to be more lithium depleted than fast rotators due to the increase in the disc lifetime. 
This relationship is particularly interesting in the context of the observation of 
higher lithium abundances for fast rotators than for slow rotators in the Pleiades \citep{sod93} and of the
possible detection of differences in the lithium content of stars with and without exoplanets reported by \cite{isr09}; but see also
\cite{bau10} for a contrasting view. The exact degree of coupling between the star 
and its disc may of course differ from the simple assumption of disc-locking used here, while the transport of angular momentum in the
radiative zone of PMS solar-type stars may be influenced by additional physical processes such as magnetic fields. We are 
indeed still far from having a complete understanding of the various physical processes at work during the evolution of solar-type stars,
but this study interestingly illustrates that the impact of rotational mixing depends on the interaction between the star and its disc.

\bibliographystyle{aa}

\bibliography{biblio}

\begin{acknowledgements}
Part of this work was supported by the Swiss National Science Foundation.
\end{acknowledgements}

\end{document}